\documentclass[apj]{emulateapj}

\usepackage{color}

\begin{document}

\title{Fermi Large Area Telescope Detection of Supernova Remnant RCW 86}

\author{Qiang Yuan$^{1,2}$, Xiaoyuan Huang$^3$, Siming Liu$^3$, 
Bing Zhang$^2$}

\affil{$^1$Key Laboratory of Particle Astrophysics, Institute of High
Energy Physics, Chinese Academy of Sciences, Beijing 100049, China\\
$^2$Department of Physics and Astronomy, University of Nevada Las Vegas,
NV 89154, USA\\
$^3$Key Laboratory of Dark Matter and Space Astronomy, Purple
Mountain Observatory, Chinese Academy of Sciences, Nanjing 210008, China}

\begin{abstract}

Using $5.4$ year Fermi-LAT data, we report the detection of GeV $\gamma$-ray 
emission from the shell-type supernova remnant RCW 86 (G315.4-2.3) with a
significance of $\sim5.1\sigma$. The data slightly favors an extended
emission of this supernova remnant. The spectral index of RCW 86 is found
to be very hard, $\Gamma\sim1.4$, in the $0.4$ to $300$ GeV range. A one
zone leptonic model can well fit the multi-wavelength data from radio to 
very high energy $\gamma$-rays. The very hard GeV $\gamma$-ray spectrum 
and the inferred low gas density seem to disfavor the hadronic origin
of the $\gamma$-rays. The $\gamma$-ray behavior of RCW 86 is very similar 
to several other TeV shell-type supernova remnants, e.g., RX J1713.7-3946, 
RX J0852.0-4622, SN 1006 and HESS J1731-347.

\end{abstract}

\keywords{radiation mechanisms: non-thermal --- gamma rays: ISM
--- ISM: supernova remnants --- cosmic rays}

\section{Introduction}

Supernova remnants (SNRs) are believed to be the most probable candidates
of the Galactic cosmic ray (CR) acceleration sources. However, direct 
observational evidence is not available until there are $\gamma$-ray 
detections of SNRs \citep[e.g.,][]{2010ApJ...710L.151T,2013Sci...339..807A}. 
Up to now, nearly $20$ SNRs have been discovered in 
TeV $\gamma$-ray band, among which $7$ are firmly identified as shell-type
SNRs \citep{2013FrPhy...8..714R} and about half are interacting with 
molecular clouds\footnote{http://tevcat.uchicago.edu}.
In the two-year catalog of Fermi-LAT (2FGL), there are $6$ firmly 
identified SNRs based on the spatial extension and $4$ associated point-like 
SNRs \citep{2012ApJS..199...31N}. Additionally there are $59$ 2FGL sources
which might be associated with SNRs based on the spatial match between the
error circles of 2FGL sources and the SNR extensions 
\citep{2012ApJS..199...31N}. With the accumulation of Fermi-LAT data,
more and more SNRs were detected \citep{2010ApJ...717..372C,
2011ApJ...740L..51T,2011ApJ...734...28A,2011ApJ...740L..12W,
2012ApJ...744L...2G,2012ApJ...744...80A,2012ApJ...752..135K,
2012ApJ...759...89H,2013ApJ...774...36C,2013MNRAS.434.2202A,
2013ApJ...779..179P,2014ApJ...781...64X,2014ApJ...783...32A}. 
Although the $\gamma$-ray emission mechanism of individual SNRs is subject 
to debate, it is possible to approach the nature of $\gamma$-ray emission 
of SNRs through a population study with a large sample of $\gamma$-ray SNRs
\citep{2012ApJ...761..133Y,2013A&A...553A..34D}. Increasing the sample
of $\gamma$-ray SNRs can be essential for understanding their
non-thermal characteristics.

The shell-type SNR G315.4-2.3, also known as RCW 86, is a young remnant 
probably associated with supernova SN 185 \citep{2002ISAA....5.....S,
2006ChJAA...6..635Z}. The angular diameter of this SNR is about $42'$, 
with a clear shell in radio 
\citep{1987A&A...183..118K,1996A&AS..118..329W,2001ApJ...546..447D}, 
infrared \citep{2011ApJ...741...96W}, optical \citep{1973ApJS...26...19V,
1997AJ....114.2664S} and X-ray bands \citep{1997A&A...328..628V,
2000A&A...360..671B,2000PASJ...52.1157B,2001ApJ...550..334B,
2002ApJ...581.1116R}. The distance of RCW 86 is
estimated to be $2.3-2.8$ kpc throught optical spectroscopy observations
\citep{1996A&A...315..243R,2003A&A...407..249S}. In the very high energy
(VHE) $\gamma$-ray band, a well extended source with morphology consistent
with the X-ray image has been revealed by HESS \citep{2009ApJ...692.1500A}.
The spectral index of VHE $\gamma$-rays is about $2.5$ and the flux is
about $10\%$ of that of the Crab nebula \citep{2009ApJ...692.1500A}.
\citet{2012A&A...545A..28L} analyzed $\sim3$ year Fermi-LAT data and
found no significant excess from this SNR. Upper limits of $\gamma$-ray
flux in the GeV band were derived \citep{2012A&A...545A..28L}. With 
multi-wavelength observations, the high energy radiation mechanism and
particle acceleration can be studied \citep{2009ApJ...692.1500A,
2012A&A...545A..28L}.

Here we report the detection of GeV $\gamma$-ray emission from RCW 86, 
with $5.4$ year Fermi-LAT data. The data analysis, including the morphology
and the spectrum, is presented in Sec. 2. Based on the $\gamma$-ray spectrum 
and the multi-wavelength spectral energy distribution (SED) of RCW 86, 
we discuss its non-thermal emission mechanism in Sec. 3. Finally Sec. 4 
is the conclusion.

\section{Data Analysis}

\begin{figure*}[!htb]
\centering
\includegraphics[width=\columnwidth]{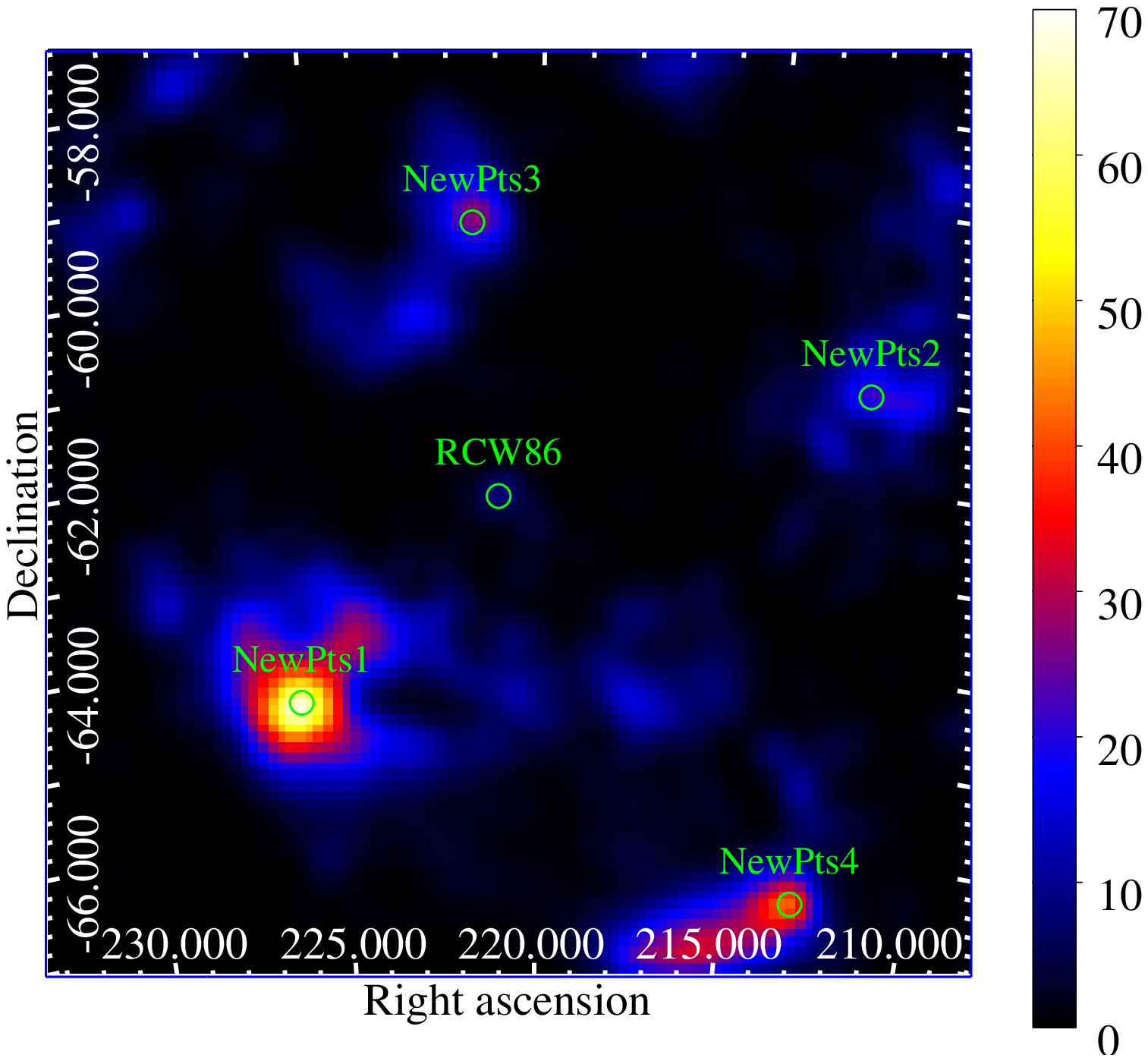}
\includegraphics[width=\columnwidth]{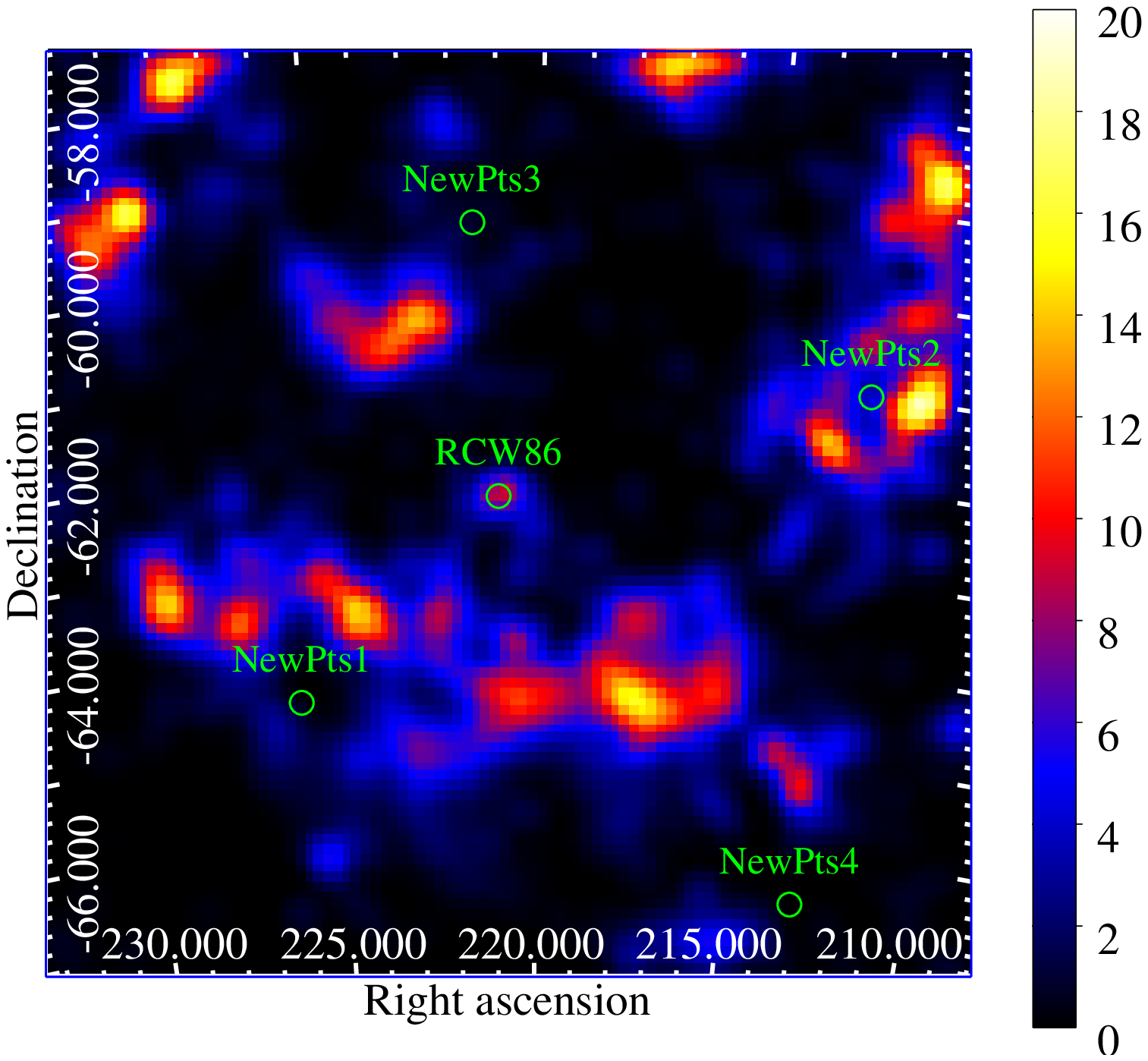}
\caption{TS maps above 400 MeV for $10^{\circ}\times10^{\circ}$ region 
centered at RCW 86. Left panel is for the model with the diffuse backgrounds 
and 2FGL sources subtracted, and the right panel is for the model with the 
additional sources listed in Table \ref{table:newpts} subtracted. Green
circles label the positions of the newly added sources and RCW 86.
The maps are smoothed with $\sigma=0.3^{\circ}$ Gaussian function.}
\label{fig:tsmap}
\end{figure*}

The newest reconstructed Pass 7 reprocessed version of the Fermi-LAT
data\footnote{http://fermi.gsfc.nasa.gov/ssc/data} are used in this analysis.
We select the data recorded from 4 August 2008 to 16 January 2014, in total 
284 weeks. The {\tt SOURCE} (evclass=2) event class is selected and the
maximum zenith angle cut is $100^{\circ}$. The data are filtered 
with the recommended cuts {\tt(DATA\_QUAL==1) \&\& (LAT\_CONFIG==1) \&\&
ABS(ROCK\_ANGLE)$<52$}. The energy range in the analysis is adopted to
be $400$ MeV to $300$ GeV, and the region-of-interest (ROI) is taken to 
be a $14^{\circ}\times14^{\circ}$ box around the position of RCW 86.
Such a box size is reasonable compared with the $\lesssim 1.5^{\circ}$
resolution angle for photons above $400$ MeV \citep{2009ApJ...697.1071A}.
The analysis is based on the LAT Scientific tool version {\tt 
v9r32p5}\footnote{http://fermi.gsfc.nasa.gov/ssc/data/analysis/software/},
and the instrument response function (IRF) is {\tt P7REP\_SOURCE\_V15}. 
The Galactic diffuse background {\tt gll\_iem\_v05.fits} and isotropic 
diffuse background {\tt iso\_source\_v05.txt} provided by the Fermi 
Science Support Center\footnote
{http://fermi.gsfc.nasa.gov/ssc/data/access/lat/Background Models.html}
are adopted in the analysis.

We bin the data into 30 logarithmically distributed energy bins and 
$140\times140$ spatial bins with size $0.1^{\circ}$, and perform the 
analysis following the standard binned likelihood analysis procedure. 
The 2FGL sources \citep{2012ApJS..199...31N} within radius $15^{\circ}$ 
around RCW 86 are included in the source model, which is generated by 
the User-contributed software {\tt 
make2FGLxml.py}\footnote{http://fermi.gsfc.nasa.gov/ssc/data/analysis/user/}.
In the likelihood fittings, the spectral parameters of all the sources
located in the ROI together with the normalizations of the two diffuse
backgrounds are left free. 

We first fit the model with only the 2FGL sources. The Test Statistic (TS,
defined as $2(\ln{\mathcal L}-\ln{\mathcal L}_0)$ with ${\mathcal L}_0$
the likelihood of null hypothesis and ${\mathcal L}$ the likelihood with
the source included) map for $10^{\circ}\times10^{\circ}$ region centered 
at RCW 86 after subtracting this baseline model is shown in the left panel 
of Fig. \ref{fig:tsmap}. The TS map is smoothed with $\sigma=0.3^{\circ}$ 
Gaussian function. From this TS map we find that there are some excesses 
which are not included in the 2FGL catalog. We will add five point sources 
close to the highest TS value locations, whose actual locations will be 
determined with the {\tt gtfindsrc} tool, to approximate such excess emission 
(see the green circles in the TS map\footnote{The initial postulated 
position of NewPts 5 is close to NewPts 4. However, the output location
from {\tt gtfindsrc} tool is out of the $10^{\circ}\times10^{\circ}$ region
of the TS map.}). At the location of RCW 86 (the center of the map) we 
see a relatively weak signal which may come from the emission from the SNR. 
We will also add RCW 86 in the new model. Power law spectra of these newly
added sources are assumed.

\begin{figure*}[!htb]
\centering
\includegraphics[width=0.33\textwidth]{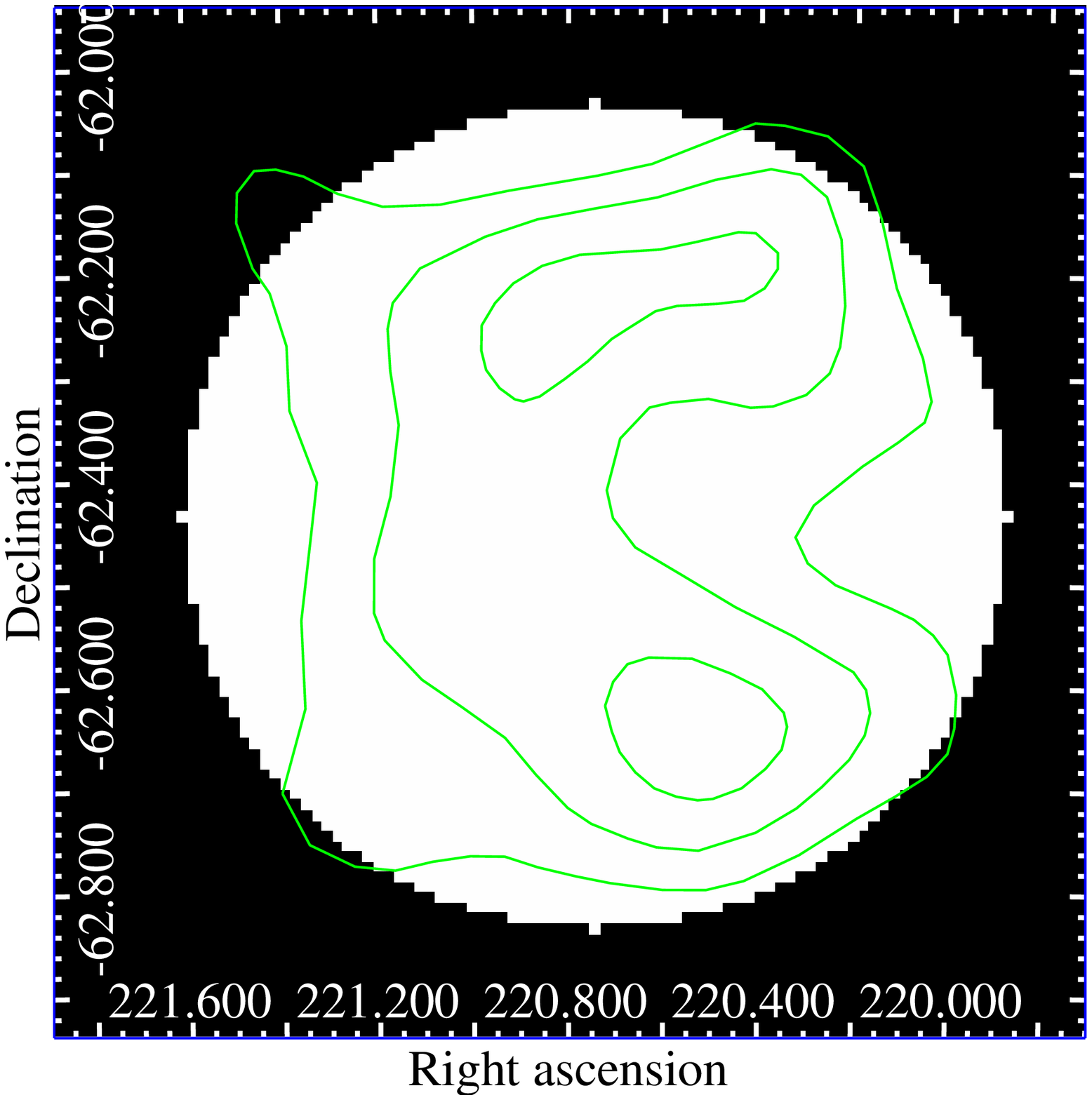}
\includegraphics[width=0.33\textwidth]{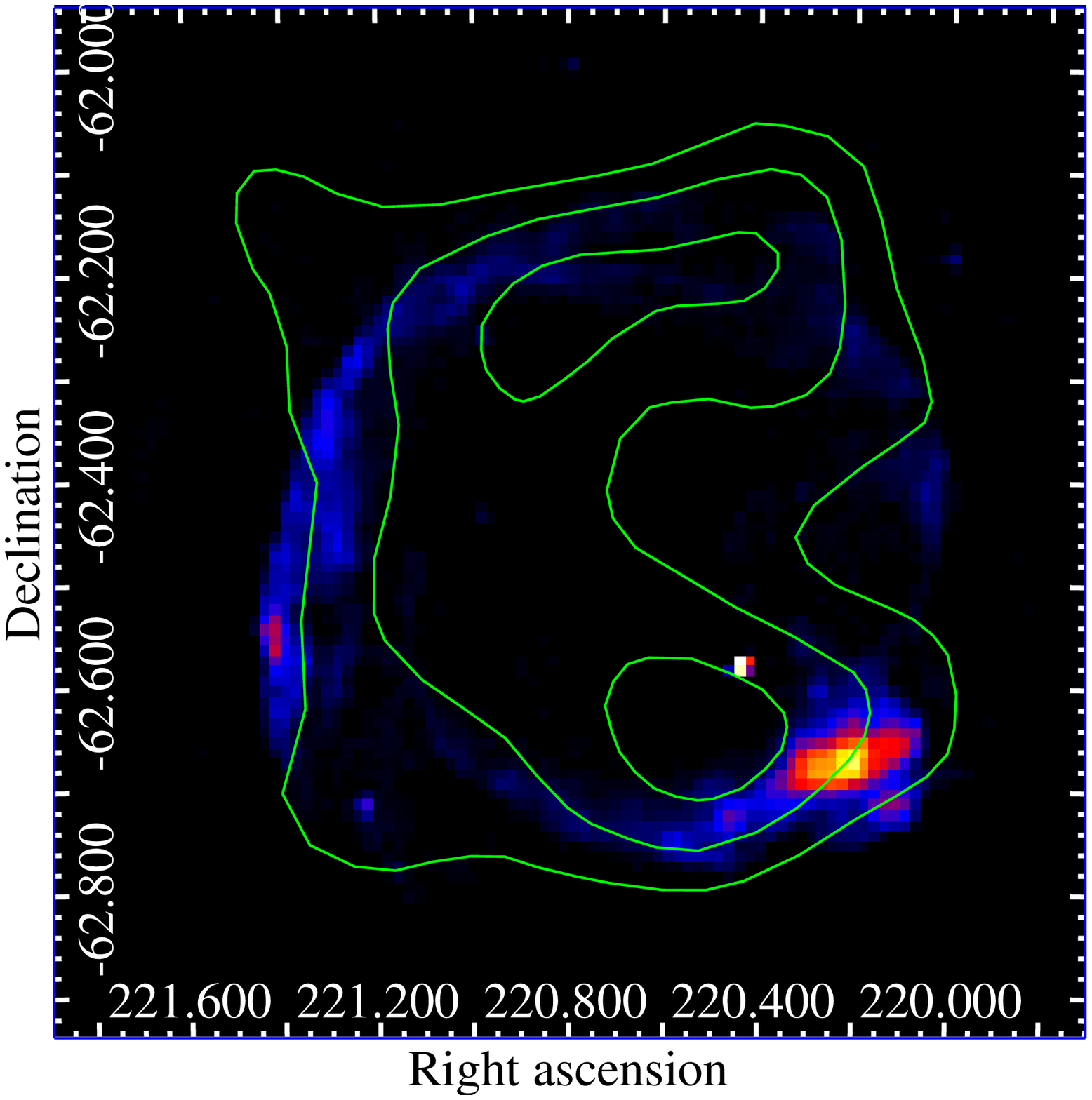}
\includegraphics[width=0.33\textwidth]{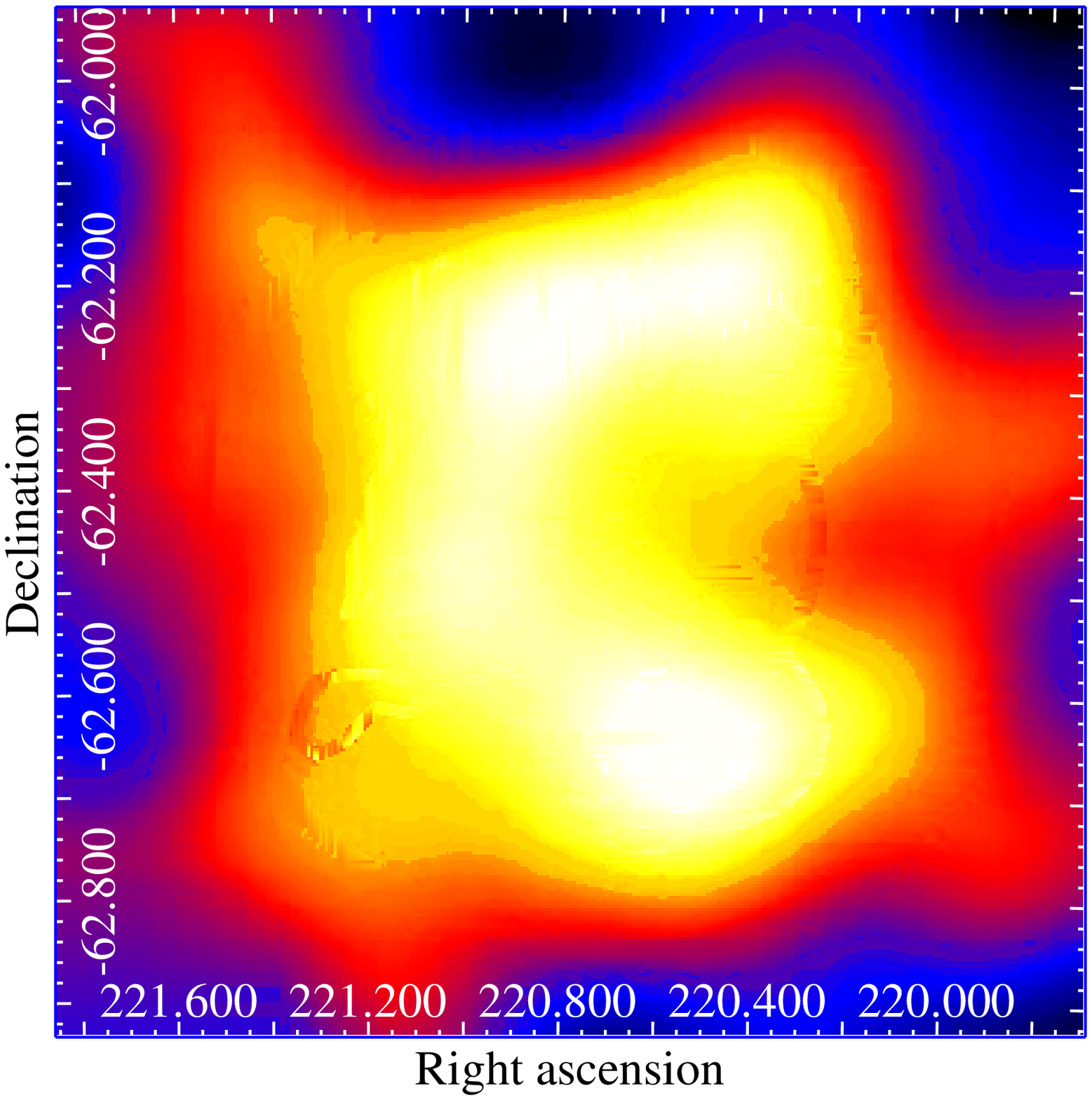}
\caption{Spatial templates used for the analysis. Left: uniform disk with
radius $0.4^{\circ}$; middle: radio image at $843$ MHz from SUMSS survey 
\citep{2003MNRAS.342.1117M}; right: HESS image \citep{2009ApJ...692.1500A}. 
Green contours overlaid on the left and middle panels are the HESS 
$\gamma$-ray excess contours \citep{2009ApJ...692.1500A}.}
\label{fig:template}
\end{figure*}

The radio and X-ray observations show clear morphology of RCW 86 
\citep{1996A&AS..118..329W,2001ApJ...546..447D,1997A&A...328..628V,
2000A&A...360..671B,2000PASJ...52.1157B,2001ApJ...550..334B,
2002ApJ...581.1116R}, and the angular radius is about $0.35^{\circ}$. 
The HESS observation of TeV $\gamma$-rays reveals the spatial 
extension with radius of about $0.4^{\circ}$ \citep{2009ApJ...692.1500A}. 
Therefore RCW 86 should be treated as an extended source in the analysis. 
We will use a uniform disk with radius $0.4^{\circ}$, the radio image at 
$843$ MHz from the Sydney University Molonglo Sky Survey (SUMSS, 
\cite{2003MNRAS.342.1117M}), and the HESS TeV $\gamma$-ray image as the
spatial template for RCW 86. The central position of the disk template
is adopted to be (R.A.$=220.75$, Dec$=-62.43$), which can well match the
HESS and SUMSS images of the SNR. The point source assumption will also be
adopted for comparison. The three extended spatial templates are shown
in Fig. \ref{fig:template}. In the left and middle panels the HESS excess
contours of $\gamma$-rays are overlaid with green lines 
\citep{2009ApJ...692.1500A}. 

With these new sources in the model, the fitting improves significantly
(the value of log-likelihood increases by $\sim 236$ compared with
the fit without these new sources).
The coordinates and TS values of the five new point sources are listed
in Table \ref{table:newpts}. The TS map after subtracting the additional
five new sources listed in Table \ref{table:newpts} is shown in the right 
panel of Fig. \ref{fig:tsmap}. It can be seen that this TS map become 
much smoother than the baseline model (the left panel). There are still
some residual excesses which might be due to the inaccuracy of the 
Galactic diffuse background or the existence of additional point sources. 
These residuals are not expected to affect the results of RCW 86 
remarkably. Actually even the most significant five new sources listed 
in Table \ref{table:newpts} are not included in the model, the fitting 
results of RCW 86 do not change significantly.

\begin{table}[!htb]
\centering
\caption {Coordinates and TS values of the new point sources.}
\begin{tabular}{cccc}
\hline \hline
Name & R.A. [deg] & Dec. [deg] & TS \\
\hline
NewPts1 & $225.91$ & $-64.46$ & $103.7$ \\
NewPts2 & $212.62$ & $-61.00$ & $48.6$ \\
NewPts3 & $221.49$ & $-59.36$ & $51.2$ \\
NewPts4 & $213.08$ & $-66.55$ & $63.6$ \\
NewPts5 & $216.22$ & $-68.15$ & $299.5$ \\
\hline
\hline
\end{tabular}
\label{table:newpts}
\end{table}

Now we focus on the discussion about RCW 86. For point source assumption
of RCW 86, the best-fit position is ${\rm RA=220.96^{\circ}}$, ${\rm Dec=
-62.32^{\circ}}$, and the TS value is $26.6$. For four degree of freedom
(dof) such a TS value corresponds to a significance $\sim4.2\sigma$. As a
comparison, in \cite{2012A&A...545A..28L} the TS value of RCW 86 for 
point source model is about $12$. We then test the three spatial templates
as shown in Fig. \ref{fig:template}. It is found that for extended source
assumption of RCW 86, the TS value is about $30$. For two dof (normalization 
and spectral index) it corresponds to a $5.1\sigma$ significance. To
better address the extension of the source, we compare the results for a
disk with very small radius (0.1 degree). The results recover the point 
source assumption. If the central position of the small disk is the same 
as the above 0.4 deg disk (the best fitting position of point source 
assumption), the TS value is about 16.9 (28.1). It shows that the data do 
favor an extended emission of the source. The fitting TS values and 
spectral indices for different spatial templates are compiled in Table 
\ref{table:result}. No significant differences among the three spatial
templates as shown in Fig. \ref{fig:template} can be found from the 
Fermi-LAT data.

\begin{table}[!htb]
\centering
\caption {Fitting results of different spatial templates.}
\begin{tabular}{ccccc}
\hline \hline
 & point & disk & SUMSS & HESS \\
\hline
TS  & $26.6$ & $32.1$ & $31.6$ & $29.6$ \\
$\Gamma$ & $1.21\pm0.32$ & $1.38\pm0.18$ & $1.36\pm0.18$ & $1.33\pm0.19$ \\
Flux$^a$ & $2.16\pm1.55$ & $6.12\pm2.95$ & $5.66\pm2.67$ & $5.37\pm2.78$ \\
\hline
\hline
\end{tabular}
$^a$Flux between $0.4$ and $300$ GeV in $10^{-10}$ cm$^{-2}$s$^{-1}$.
\label{table:result}
\end{table}

\begin{figure*}[!htb]
\centering
\includegraphics[width=\columnwidth]{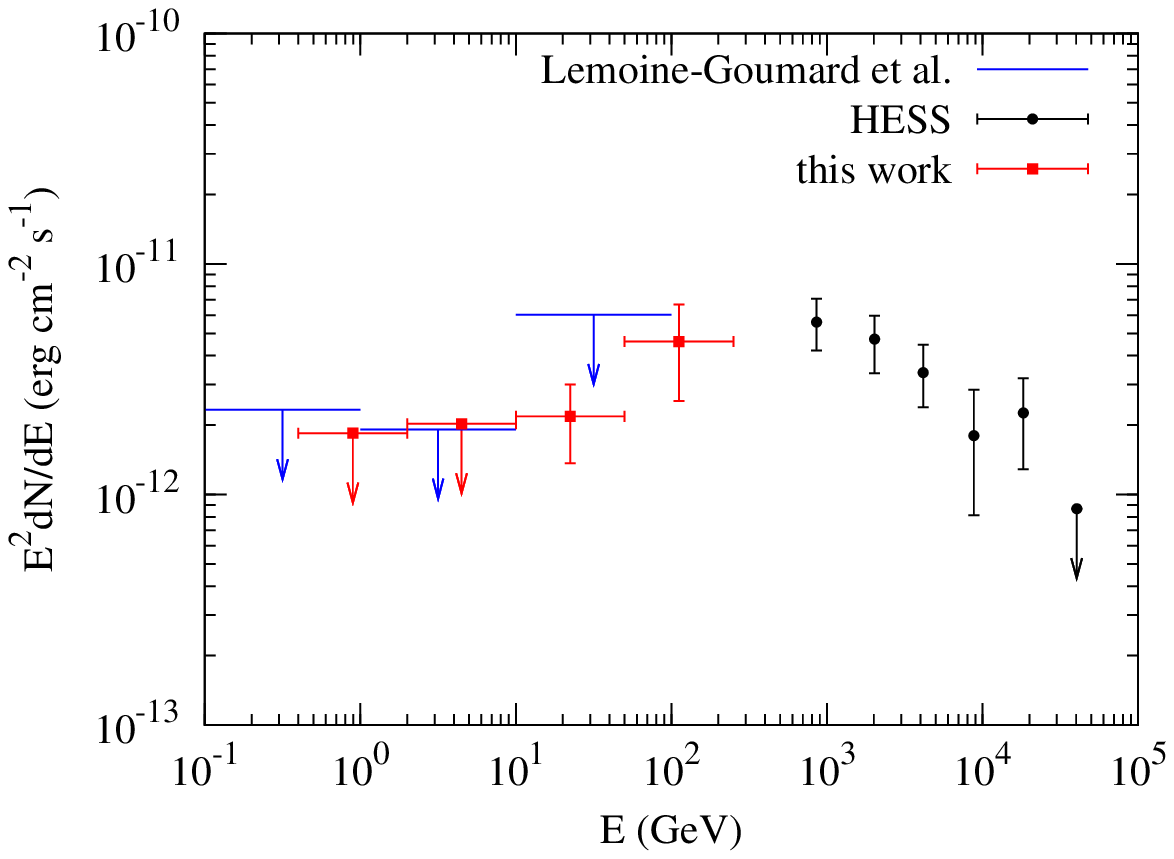}
\includegraphics[width=\columnwidth]{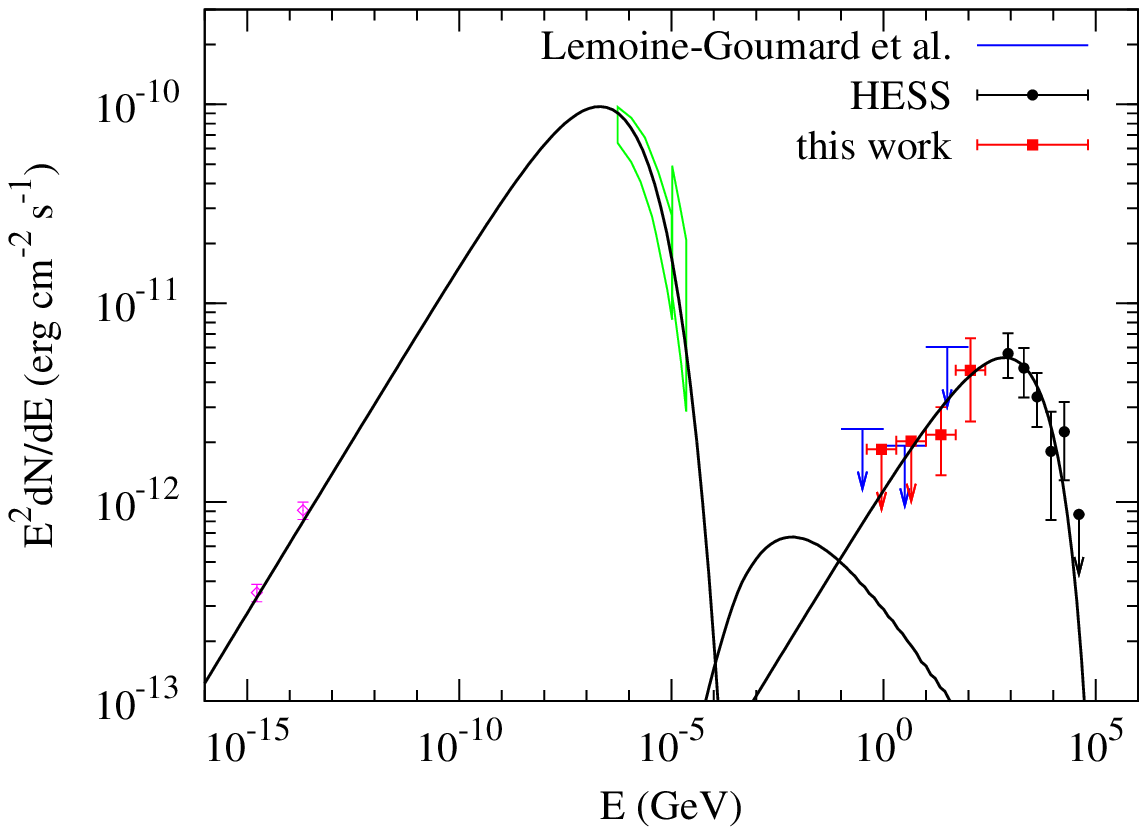}
\caption{Left: $\gamma$-ray spectrum of RCW 86 for the disk template. 
Black dots are the HESS data in the VHE band \citep{2009ApJ...692.1500A}, 
and blue arrows are the upper limits derived using 40 month Fermi data 
\citep{2012A&A...545A..28L}. Right: multi-wavelength SED of RCW 86.
Radio data are from Molongo at 408 MHz and Parkes at 5 GHz 
\citep{1975AuJPA..37...39C}; X-ray data from ASCA and RXTE 
\citep{2012A&A...545A..28L}. Lines show a one-zone leptonic modeling of
the data, with the three bumps from left to right synchrotron, 
bremsstrahlung and inverse Compton scattering components respectively.}
\label{fig:spec}
\end{figure*}

We test different central positions of the disk template through 
increasing or decreasing the R.A. or Dec. by 0.1 degree. The resulting 
TS values of RCW 86 decrease by $0.7-4.1$. We also test the disk 
templates with 0.3 and 0.5 degree radii, and get the TS value 30.4 and 
29.6 respectively, which are also slightly smaller than the value 32.1 
as given in Table \ref{table:result}. It is shown that the disk template
shown in Fig. \ref{fig:template} does fit the data well. The best-fitting 
spectral indices in these tests differ by about 0.04, which could be 
regarded as systematics due to the choice of position and extension of 
the disk.

The spectrum for RCW 86 is very hard. Such a hard spectrum will be 
difficult to be explained with the hadronic scenario whose $\gamma$-ray
spectrum just follows the proton spectrum. The inverse Compton scattering 
of background photons by high energy electrons, on the other hand, can 
easily account for the hard spectrum revealed by the Fermi-LAT data. 
Similar hard spectra of GeV photons are also shown in two other shell-type 
SNRs RX J1713.7-3946 \citep{2011ApJ...734...28A} and RX J0852.0-4622 
\citep{2011ApJ...740L..51T}. For another two shell-type SNRs SN 1006 
\citep{2012MNRAS.425.2810A} and HESS J1731-347 \citep{Yang_HESSJ1731}, 
although they have not been detected yet in Fermi-LAT data, the flux 
upper limits actually show similar behaviors for the GeV-TeV spectra 
like RCW 86.

We derive the SED of RCW 86 with the same likelihood analysis, but done 
in different energy bins. The spectral indices of all the sources are 
fixed to be the best-fit values obtained in the previous global fitting, 
only the normalizations of the sources and the diffuse backgrounds are 
free during the fittings. The SED for the disk template is shown in 
the left panel of Fig. \ref{fig:spec}. For the other two extended source 
templates the results are essentially similar. In the first two energy bins 
the TS values for RCW 86 are very small, and the $99.9\%$ upper limits are 
given. For comparison the upper limits obtained in \cite{2012A&A...545A..28L} 
and the HESS data in the VHE band \citep{2009ApJ...692.1500A} are also 
shown. The GeV $\gamma$-ray SED derived in this work is consistent with 
the upper limits obtained in \cite{2012A&A...545A..28L}. It can be seen 
that connecting the GeV-TeV SED shows a peak at hundreds of GeV. Such a 
peak may indicate the leptonic feature of the $\gamma$-ray emission.

\section{Discussion}

In the right panel of Fig. \ref{fig:spec} we compile the multi-wavelength 
observational data of RCW 86, from radio \citep{1975AuJPA..37...39C}, X-ray 
\citep{2012A&A...545A..28L}, to VHE $\gamma$-rays \citep{2009ApJ...692.1500A}.
This wide band SED shows a double-peak behavior, which might be reasonably
described within the leptonic framework. The high energy electron spectrum
is parameterized with an exponential cutoff power-law spectrum, $dN/dE_e
\propto E_e^{-\Gamma_e}\exp(-E_e/E_c)$. The total energy of electrons above 
1 GeV is normalized to $W_e$, which could be a fraction of the total energy
released by the supernova. We consider a simple one zone model, where a 
single electron population radiates in a uniform magnetic field and matter 
field. For the background photons used to calculate the inverse
Compton scattering emission, we adopt the interstellar radiation field
(ISRF) as developed in \cite{2005ICRC....4...77P}, which is made up of
optical emission from starlight, infrared from absorption and re-emission
from dust and the cosmic microwave background (CMB). To calculate the
bremsstrahlung emission, we also adopt a gas number density of $1$ cm$^{-3}$. 
Adopting proper parameters, we reproduce the multi-wavelength emission 
of RCW 86 well, as shown in the right panel of Fig. \ref{fig:spec}. 
The three bumps from left to right represent the synchrotron, bremsstrahlung 
and inverse Compton scattering components generated by the same population 
of electrons. Although the radio and X-ray images show complex structures,
the gamma-ray observations are fully consistent with the one-zone model. 
The X-ray emission has contributions from a thermal component, and the 
synchrotron emission is also affected by the magnetic field structure. 
There is no compelling evidence for a two-zone emission model.

Assuming that the distance of RCW 86 is $2.5$ kpc, and the radius is $15$ pc,
we estimate the model parameters to reproduce the multi-wavelength SED, which
are $\Gamma_e\approx2.3$, $E_c\approx22$ TeV, $W_e\approx1.2\times10^{48}$
erg, and $B\approx20$ $\mu$G. These parameters are consistent with\footnote{
Except for $W_e$ which might be due to different low threshold energy to 
calculate the total electron energy.} that of the one zone model in 
\cite{2012A&A...545A..28L}. The total energy goes to high energy electrons
is about $0.1\%$ of the typical released energy of a supernova, say $10^{51}$
erg. We also test the case with only the CMB as the target photon field
to produce the inverse Compton scattering emission. In this case, the
total energy of electrons ($W_e$) needs to be $\sim2$ times higher, the 
magnetic field is about $\sqrt{2}$ times smaller, and the cutoff energy 
$E_c\approx26$ TeV which is slightly larger. The resulting photon spectrum 
for the CMB photon field case is very similar as the one shown in the 
right panel of Fig. \ref{fig:spec}.

It should be noted that the cooling of electrons is not effective enough
to alter the electron spectrum of RCW 86. From Fig. \ref{fig:spec} we
see that most of the electron energy goes into synchrotron radiation.
The synchrotron cooling time is estimated as $\tau_{\rm syn}\approx
1.25\times 10^{10}(B/\mu{\rm G})^{-2}(E/{\rm GeV})^{-1}$ yr. The age
of RCW 86 is about $1800$ yr, thus the critical energy above which
the electrons are cooled down is about 35 (17) TeV for magnetic field
14 (20) $\mu$G. Since our cutoff energy is about $20$ TeV, the cooling
does not affect the electron spectrum assumed in the modeling.

The hadronic scenario seems not favored to explain the $\gamma$-ray
spectrum since it requires too hard a spectrum of the accelerated
cosmic ray protons, which can not be easily understood in the shock
acceleration theory. There might be another problem for the hadronic 
scenario, which is the low ambient medium density as revealed by the 
thermal X-ray emission. In the above calculation we arbitrarily adopt 
a gas density of $1$ cm$^{-3}$. From the weak thermal X-ray emission, 
the post-shock density of this SNR was estimated to be $(0.26-0.68)
f^{-0.5}$ cm$^{-3}$, where $f$ is the filling factor of the thermal 
component \citep{2008PASJ...60S.123Y}. A low density is also expected
according to the high shock velocity of this SNR \citep{2009Sci...325..719H}.
Given a low density, the required energy for proton acceleration will 
be too high. For example, for the proton spectrum with index $1.7$ 
and cutoff energy $50$ TeV, the estimated total energy above 1 GeV is
$1.3\times10^{50}\left(n_{\rm gas}/1\,{\rm cm}^{-3}\right)^{-1}\left(
d/{2.5\,{\rm kpc}}\right)^2$ erg. The energy fraction of cosmic rays
in the total energy released by the supernova may be too large if the 
density is lower than $1$ cm$^{-3}$. In a recent study, 
\cite{2014A&A...562A.141M} inferred the cosmic ray
acceleration efficiency would be $\sim20\%$ based on the Balmer line
emission. Therefore the energy budget will be challenged in the hadronic
scenario. Similar circumstances also appear for SNR RX J1713.7-3946 
\citep{2010ApJ...712..287E,2011ApJ...735..120Y} and RX J0852.0-4622 
\citep{2011ApJ...740L..51T}, both having very hard GeV $\gamma$-ray spectra 
and no detection of thermal X-rays. Note \cite{2011ApJ...731...32F}
proposed the hadronic scenario based on non linear diffusive shock
acceleration \citep{2001RPPh...64..429M,2002APh....16..429B} to explain 
the $\gamma$-ray emission of these three SNRs. The model prediction 
is generally higher and harder than the Fermi-LAT observations.

The GeV-TeV $\gamma$-ray emission of this SNR, together with the fact that
the ambient medium density may be low, further supports the unified picture 
to describe the $\gamma$-ray SNRs \citep{2012ApJ...761..133Y}. In that
model, the SNRs are classified into three classes according to the medium
density. The sources located in low density medium tend to have an inverse
Compton scattering origin of $\gamma$-rays, and the $\gamma$-ray spectrum
will be very hard. Up to now, there are probably five shell-type SNRs
belong to this class, i.e., RX J1713.7-0846, RX J0852.0-4622, RCW 86,
SN 1006 and HESS J1731-347. A combined study of all these sources and an 
evolutionary picture to describe them will be interesting.

\section{Conclusion}

In this work we report the detection of GeV $\gamma$-rays from a shell-type 
SNR RCW 86 with Fermi-LAT. Analyzing $5.4$ year Fermi-LAT data, we find
an extended source coincident with the radio or VHE $\gamma$-ray image
of RCW 86 with a significance higher than $5\sigma$. The point source 
assumption is less favored than the extended source assumption. The GeV
$\gamma$-ray spectrum is found to be very hard, $\Gamma\approx1.4\pm0.2$.
The multi-wavelength SED from radio to VHE $\gamma$-rays can be well
described with a simple one zone leptonic model. The hadronic scenario
may face difficulty in producing the very hard GeV $\gamma$-ray spectrum 
and in the total energy budget given that the environmental medium density
is relatively low. The analogy of the non-thermal GeV-TeV spectrum and 
the lack of strong thermal X-ray emission of this SNR with several other 
shell-type SNRs makes them form a distinct class of $\gamma$-ray SNRs, 
which may point to the nature of the particle acceleration and 
radiation in SNRs.

\section*{Acknowledgements}
We acknowledge the use of the Fermi-LAT data provided by the Fermi
Science Support Center, and the anonymous referee for useful comments
and suggestions. QY thanks Ye Li for helpful discussion. 
This work is supported by 973 Program under Grant No. 2013CB837000, 
the National Natural Science Foundation of China under Grant Nos. 
11105155, 11173064 and 11233001, and the Strategic Priority Research 
Program -- The Emergence of Cosmological Structures of the Chinese 
Academy of Sciences under Grant No. XDB09000000.


\end{document}